\documentclass[preprint,12pt]{elsarticle}
\usepackage{psfrag}
\usepackage{graphicx}
\usepackage{paralist}
\usepackage{algpseudocode}
\usepackage{color}
\usepackage{eucal}
\usepackage{amsmath}
\usepackage{amssymb}
\usepackage{amsthm}
\usepackage{varioref}
\usepackage{fancyhdr}
\usepackage{amsfonts}
\usepackage{bm}
\usepackage{float}
\usepackage{booktabs}
\usepackage{multirow}
\usepackage{array}
\usepackage{algorithm}
\usepackage{mathptmx}
\usepackage{natbib}
\usepackage{lineno}
\usepackage{epstopdf}
\usepackage{microtype}
\usepackage{tcolorbox}
\usepackage[tikz]{bclogo}
\usepackage{lipsum}
\usepackage[breaklinks,colorlinks,linkcolor=blue,citecolor=blue,urlcolor=black]{hyperref}
\usepackage{caption}
\usepackage{threeparttable}
\newcommand{\bfg}{\boldsymbol}
\newcommand{\real}{\hbox{\rm I\kern-0.2emR}}
\newtheorem{remark}{Remark}[section]
\newtheorem{theorem}{Theorem}[section]
\newtheorem{corollary}{Corollary}[theorem]

\journal{arXiv.org}
\begin{document}
\begin{frontmatter}
\title{Peridynamic stress is a weighted static Virial stress}

\author[label1]{Shaofan Li\corref{cor1}}

\address[label1]{Department of Civil and Environmental Engineering, University of California, Berkeley, CA 94720, USA}

\cortext[cor1]{Corresponding author: shaofan@berkeley.edu}

\begin{abstract}
The peridynamic stress tensor proposed by Lehoucq and Silling \cite{Lehoucq2008}
is cumbersome to implement in numerical computations.
In this note, we show that the peridynamic stress tensor has
the mathematical expression 
of a weighted static Virial stress derived by Irving and Kirkwood \cite{Irving1950},
which offers a simple and clear expression for numerical calculations
of peridynamic stress tensor.
\end{abstract}

%\begin{keyword}
%Irving-Kirkwoods formalisim \sep
%Nonlocal continuum mechanics \sep
%Nonlocal differential operator \sep
%Peridynamics \sep
%Peridynamic stress tensor \sep
%Virial stress \sep
%\end{keyword}
\end{frontmatter}

%\linenumbers

\bigskip
\bigskip

Peridynamics is reformulation of non-local continuum mechanics
or computational non-local mechanics \cite{Silling2000,Silling2010}.
In the peridynamic equation of motion,
following the notation of Silling and Lehoucq \cite{Silling2010}
one can write the balance of linear momentum as
\begin{equation}
\rho \ddot{\bf u} ({\bf X}, t) = \int_{\mathcal{H}_X}
 {\bf f}({\bf X}^{\prime}, {\bf X}, t)  d V_{X^{\prime}} + {\bf b}({\bf X},t),~~\forall {\bf X}
 \in \mathcal{B}
\label{eq:Nonlocal-BLM1}
\end{equation}
where $\mathcal{B} \subset \real^3$;  $\mathcal{H}_X \subset \mathcal{B}$ is
a the horizon of the material point ${\bf X}$;
 $\rho$ is the material density, and ${\bf b} ({\bf X}, t)$ is the body force.
The term
\[
 \int_{\mathcal{H}_X} {\bf f}({\bf X}^{\prime}, {\bf X}, t)  d V_{X^{\prime}}
\]
replaces the divergence of the first Piola-Kirchhoff stress $\nabla \cdot {\bf P}$
at the material point ${\bf X}$.

By definition,
\begin{equation}
{\bf f} ({\bf X}^{\prime}, {\bf X}, t) :=
\bigl(
{\bf t}({\bf  X}^{\prime}, {\bf X}, t) - {\bf t}({\bf X}, {\bf X}^{\prime}, t) \bigr)
= - {\bf f} ({\bf X}, {\bf X}^{\prime}, t)
\label{eq:Fdef}
\end{equation}
is antisymmetric,
where
${\bf X}, {\bf X}^{\prime}$ are the position vectors
of material points in the referential configuration.

Equation (\ref{eq:Nonlocal-BLM1}) extends
the balance equation of linear momentum to nonlocal media.
However, it loses some valuable properties that
are associated with the local balance law such as
the divergence theorem or the Gauss theorem.

Noticing such inadequacy,
Lehoucq and Silling \cite{Lehoucq2008} define the following
nonlocal {\it Peridynamic Stress} tensor
\begin{equation}
{\bfg \varsigma} ({\bf X}) :=  {1 \over 2} \int_{\mathcal{S}^2}
\int_{0}^{\infty}
\int_{0}^{\infty} (y+z)^2
 {\bf f}({\bf X} +y {\bf M}, {\bf X} - z {\bf M}) \otimes
 {\bf M} dz d y d \Omega_M~,
\label{eq:Nonlocal-BLM2}
\end{equation}
where $\mathcal{S}^2$ is the unit sphere.
By doing so, we have the relation
\begin{equation}
\nabla \cdot {\bfg \varsigma} = \int_{\mathcal{H}_X} {\bf f}({\bf X}^{\prime},
{\bf X}, t) dV_{X^{\prime}}~.
\end{equation}
where $\nabla$ is the local gradient operator.
An immediate benefit of Eq. (\ref{eq:Nonlocal-BLM2}) is
that we can link the divergence of the peridynamic stress with
the boundary linear momentum flux, i.e.
\[
\int_{\mathcal{B}} \nabla \cdot {\bfg \varsigma} dV_{X} = \int_{\partial \mathcal{B}}
{\bfg \varsigma} \cdot {\bf N} d S_{X}~,
\]
which allows us to establish peridynamics-based Galerkin weak formulations conveniently,
and maybe even formulate peridynamics theories of plates and shells.

By using Noll's lemmas \cite{Noll1955},
such nonlocal integral theorems have been late extended to a more general situations
by Gunzburger and Lehoucq \cite{Gunzburger2010} and Du et. al.
\cite{Du2013}.
However, in practice the peridynamic stress defined in Eq. (\ref{eq:Nonlocal-BLM2})
is cumbersome to evaluate. To resolve this issue,
in this note, we show that in peridynamic particle formulation, which is
a special case of the non-local continuum,
the peridynamic stress tensor has the exact expression of the static Virial stress
defined by Irving and Kirkwood \cite{Irving1950}.

\begin{theorem}[Alternative form of {\it Peridynamic Stress} tensor]
~~{}\\
\smallskip
Consider the peridynamics force density that can be expressed as the following expression of
the Irving-Kirkwoods formulation
\cite{Irving1950,Silling2010},
\begin{equation}
{\bf f}({\bf X}^{\prime}, {\bf X}, t)
= \sum_{I=1}^{N_X} \sum_{J=1}^{N_X} {\bf F}_{IJ} \delta ({\bf X} - {\bf X}_I) \delta ({\bf X}^{\prime}
-{\bf X}_J)
\end{equation}
where ${\bf F}_{IJ}$ is the force acting on the particle $I$ from the particle $J$ (see Fig. 1);
$N_X$ is the total number of particles inside the horizon $\mathcal{H}_X$.
The nonlocal peridynamic stress
\begin{equation}
{\bfg \varsigma} ({\bf X}) :=  {1 \over 2} \int_{\mathcal{S}^2}
\int_{0}^{\infty}
\int_{0}^{\infty} (y+z)^2
 {\bf f}({\bf X} +y {\bf M}, {\bf X} - z {\bf M}) \otimes
 {\bf M} dz d y d \Omega_M~,
\label{eq:Nonlocal-ST1}
\end{equation}
has the following exact form,
\begin{equation}
{\bfg \varsigma} ({\bf X}) := -
{1 \over 2} \sum_{I=1}^{N_X} \sum_{J =1}^{N_X} {\bf F}_{IJ} \otimes
{({\bf X}_I - {\bf X}) 
\over |{\bf X}_I - {\bf X}|}|{\bf X}_I -{\bf X}_J|,~~{\bf X}_J, {\bf X}_I \in
\mathcal{H}_X,~~{\bf X}_J \not = {\bf X}_I~.
\label{eq:Nonlocal-ST2}
\end{equation}
where ${\bf F}_{IJ}$ is the force acting on the atom $I$ by the atom $J$.
\end{theorem}
\begin{figure}[H]
\centering
\includegraphics[width=4.5in]{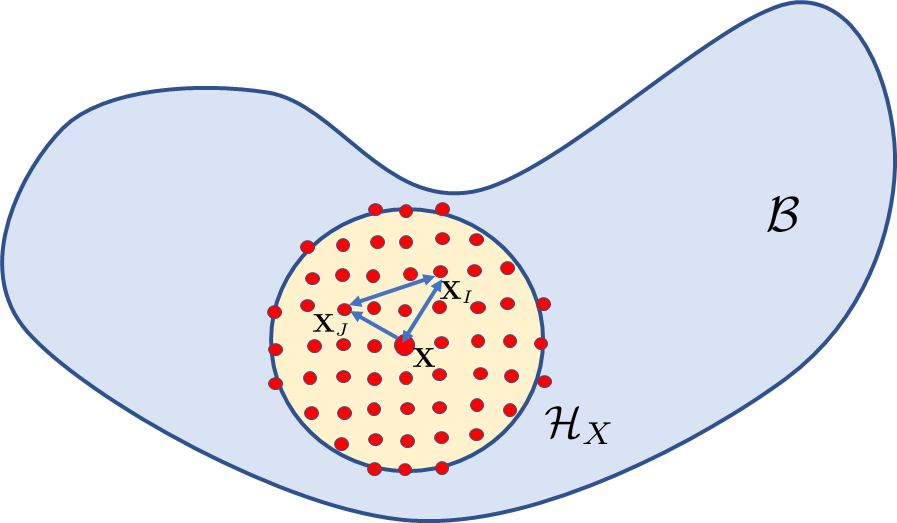}
\caption{
Illustration of peridynamics particle sampling strategy
}
\label{fig:fig1}
\end{figure}
\begin{proof}
$~{}$\\
Based on Noll's lemma \cite{Noll1955}, we can write the
first peridynamics Piola-Kirchhoff stress as
\begin{eqnarray}
{\bfg \varsigma} ({\bf X}) &=& {1 \over 2} \int_{S^2} d \Omega_m
\int_0^{\infty} R^2 d R \int_0^1 {\bf f} ({\bf X} + \alpha R {\bf M},
{\bf X} - (1-\alpha) R {\bf M}) \otimes {\bf M} d \alpha
\nonumber
\\
&=&- {1 \over 2} \int_{\real^3} d V_R
\int_0^1 {\bf f} ({\bf X} + \alpha  {\bf R},
{\bf X} - (1-\alpha)  {\bf R}) \otimes {\bf R} d \alpha~,~\forall {\bf X} \in \mathcal{B}
\label{eq:Ntilde2}
\end{eqnarray}

Considering the Irving-Kirkwood formalism \cite{Irving1950},
we have the following peridynamics sampling formulation (see Fig. 1)
\begin{equation}
{\bf f} ({\bf X}^{\prime}, {\bf X}) = - {\bf f}({\bf X}, {\bf X}^{\prime}) =-\sum_{J=1}^N{\bf F}_{IJ}
\delta ({\bf X} - {\bf X}_I)
\delta ({\bf X}^{\prime} - {\bf X}_J),~~{\bf X}_I \not = {\bf X}_J~.
\label{eq:Irving1}
\end{equation}
Letting
\[
{\bf X} = {\bf X}_C + \alpha {\bf R}, ~{\rm and}~~ {\bf X}^{\prime} = {\bf X}_C -(1-\alpha) {\bf R}
\]
and substituting them into Eq. (\ref{eq:Irving1}), we then have
\begin{eqnarray}
&&{\bf f} ({\bf X} +\alpha {\bf R}, {\bf X} - (1-\alpha) {\bf R})
\nonumber
\\
&=& \sum_{I=1}^{N_X}\sum_{J=1}^{N_X} {\bf F}_{IJ}
\delta (\alpha{\bf R} - ({\bf X}_I - {\bf X}_C))
\delta (\alpha {\bf R} -{\bf R} - ({\bf X}_J-{\bf X}_C)).
\label{eq:Irving2}
\end{eqnarray}
where ${\bf X}_I, {\bf X}_J \in \mathcal{H}_{X_C},~~{\bf X}_I \not = {\bf X}_J$.

Considering the following integration identities
\begin{eqnarray}
&& \delta (g(x)) = {\delta (x-x_0) \over |g^{\prime} (x_0)|}
\\
&&\int_{-\infty}^{\infty} \delta (\xi - x) \delta (x-\eta) d x = \delta(\eta -\xi)
\end{eqnarray}
we have
\begin{eqnarray}
&&\int_{\real^3}
\delta (\alpha{\bf R} - ({\bf X}_I - {\bf X}_C))
\delta (\alpha {\bf R} -{\bf R} - ({\bf X}_J-{\bf X}_C)) {\bf R} d (\alpha V_R)
\nonumber
\\
&&=  \alpha^{-1} ({\bf X}_I - {\bf X}_C )
\delta (({\bf X}_I-{\bf X}_J) - \alpha^{-1} ({\bf X}_I - {\bf X}_C ))~.
\end{eqnarray}
Let $\xi = \alpha^{-1}$ and $d \xi = -{ \xi \over \alpha} d \alpha \to 
d \alpha = - \xi^{-2} d \xi$.
We then can have,
\begin{eqnarray}
{\bfg \varsigma} ({\bf X}_C) &=& -{1 \over 2}\bigl(
\sum_{I=1}^{N_X}\sum_{J=1}^{N_X} {\bf F}_{IJ}
\bigr) \otimes ({\bf X}_I - {\bf X}_C) 
\nonumber
\\
&&\cdot  \int_{0}^{\infty}
\delta (({\bf X}_I-{\bf X}_J) - \xi ({\bf X}_I - {\bf X}_C )) \xi^{-1} d \xi
\nonumber
\\
&=&
-{1 \over 2} 
\sum_{I=1}^{N_X}\sum_{J=1}^{N_X} {\bf F}_{IJ}
\otimes \Bigl( {{\bf X}_I - {\bf X}_C 
\over |{\bf X}_I - {\bf X}_C| 
} \Bigr) |{\bf X}_I - {\bf X}_J| 
\label{eq:Ntilde3}
\end{eqnarray}
Let ${\bf X}_C \to {\bf X}$. We prove the claim.
\end{proof}

\begin{remark}
The above expression may be rewritten as
\begin{eqnarray}
{\bfg \varsigma} ({\bf X}) &=&
-{1 \over 2} 
\sum_{I=1}^{N_X}\sum_{J=1}^{N_X} {\bf F}_{IJ}
\otimes ({\bf X}_I - {\bf X}_J) \Bigl( {|{\bf X}_I - {\bf X}_J|
\over {\bf X}_I - {\bf X}_J
} \Bigr) \cdot
 \Bigl( {{\bf X}_I - {\bf X}
\over |{\bf X}_I - {\bf X}|
} \Bigr)~.
\label{eq:Ntilde3}
\end{eqnarray}
which is a weighted static Virial stress.
\end{remark}

For the state-based peridynamics, we have the following result.
\begin{corollary}[Peridynamics PK-I stress tensor]
~~{}\\
\smallskip
For continuous variable ${\bf X}_i, {\bf X}_j \in \mathcal{H}_C$, we let
\begin{equation}
 {\bf f} ({\bf X}_i, {\bf X}_j) =w(X_{ij})
\Bigl(
{\bf P}_j {\bf K}_j {\bf X}_{ij} - {\bf P}_i {\bf K}_i {\bf X}_{ji}
\Bigr)
\end{equation}
where ${\bf K}_i = {\bf K} ({\bf K}_i)$ is the shape tensor; $w(X_{ij} = w(|{\bf X}_j-{\bf X}_i|)$
is a window function;
${\bf X}_{ij} ={\bf X}_j - {\bf X}_i$,
and ${\bf P}({\bf X}_i)$ is the first Piola-Kirchhoff stress at the local continuum material
point ${\bf X}_i$ .
For the first Piola-Kirchhoff stress tensor, ${\bf P} ({\bf X}),~{\bf X} \in \Omega_0$,
the nonlocal divergence of ${\bf P}$,
\begin{equation}
\mathfrak{D} [{\bf P}]=
\int_{\Omega_0} w(X_{ij})
\Bigl(
{\bf P}_j {\bf X}_{ij} - {\bf P}_i {\bf X}_{ji}
\Bigr) \cdot {\bf K}^{-1} d V_j,~~\forall {\bf X}_i \in \Omega_0,
\end{equation}
equals to the local divergence of a nonlocal counterpart, $\widetilde{\bf P}$,
\begin{equation}
\nabla \cdot \widetilde{\bf P}= \mathfrak{D}  [{\bf P}]
\label{eq:LS-theorem2}
\end{equation}
where
\begin{eqnarray}
\widetilde{\bf P} ({\bf X}) &:=&
{1 \over 2} \sum_{I=1}^{N_X} \sum_{J =1, J \not=I }^{N_X}
{\bf f} ({\bf X}_I, {\bf X}_J) \otimes ({\bf X}_I - {\bf X}_J) 
\Bigl(
{|{\bf X}_I - {\bf X}_J | \over ({\bf X}_I - {\bf X}_J)}
{|{\bf X}_I - {\bf X} | \over ({\bf X}_I - {\bf X})},~~
\nonumber
\\
&&
\forall~ 
{\bf X} \in \mathcal{B},~~{\bf X}_I , {\bf X}_J \in \mathcal{H}_X~.
\label{eq:Ptilde2}
\end{eqnarray}
\end{corollary}

\begin{remark}
Note that the above result does not restricted to peridynamics PK-I stress tensor,
and it is valid for general peridynamics stress tensor after small modifications
depending on the case that is under consideration.
Amazingly, the result reveals that fact that
the mesoscale peridynamic stress tensor has exactly the same expression
as that of the microscale static Virial stress,
except that it does not count for the contribution from the kinetic
energy. Moreover, the expression (\ref{eq:Nonlocal-ST2}) is so simple
that it can be readily implemented in numerical calculations without much
trouble.
\end{remark}

\section*{Reference}

\bibliographystyle{unsrt}
\bibliography{ref}

\begin{thebibliography}{1}

\bibitem{Lehoucq2008}
R.B. Lehoucq and S.A. Silling.
\newblock Force flux and the peridynamic stress tensor.
\newblock {\em Journal of the Mechanics and Physics of Solids},
  56(4):1566--1577, 2008.

\bibitem{Irving1950}
J.H. Irving and J.G. Kirkwood.
\newblock The statistical mechanical theory of transport processes. iv. the
  equations of hydrodynamics.
\newblock {\em The Journal of Chemical Physics}, 18(6):817--829, 1950.

\bibitem{Silling2000}
S.A. Silling.
\newblock Reformulation of elasticity theory for discontinuities and long-range
  forces.
\newblock {\em Journal of the Mechanics and Physics of Solids}, 48(1):175--209,
  2000.

\bibitem{Silling2010}
S.A. Silling and R.B. Lehoucq.
\newblock Peridynamic theory of solid mechanics.
\newblock {\em Advances in Applied Mechanics}, pages 73--168, 2010.

\bibitem{Noll1955}
W.~Noll.
\newblock Die herleitung der grundgleichungen der thermomechanik der kontinua
  aus der statistischen mechanik.
\newblock {\em Journal of Rational Mechanics and Analysis}, 4:627--646, 1955.

\bibitem{Gunzburger2010}
M.~Gunzburger and R.B. Lehoucq.
\newblock A nonlocal vector calculus with application to nonlocal boundary
  value problems.
\newblock {\em Multiscale Modeling \& Simulation}, 8(5):1581--1598, 2010.

\bibitem{Du2013}
Q.~Du, M.~Gunzburger, R.B. Lehoucq, and K.~Zhou.
\newblock A nonlocal vector calculus, nonlocal volume-constrained problems, and
  nonlocal balance laws.
\newblock {\em Mathematical Models and Methods in Applied Sciences},
  23(03):493--540, 2013.

\end{thebibliography}

\end{document}